\documentclass[12pt]{article}
\usepackage{amsfonts}
\usepackage{amsmath}
\usepackage{amsthm}
\usepackage{fancybox}
\DeclareMathOperator*{\wlim}{w-lim}
\newcommand{\ps}[1]{\sigma (#1)} 
\newcommand{\latt}[2]{\|#1,#2\rangle \!\rangle}
\newcommand{\cantor}{(\mathbb{Z}_{2})^{\omega}}

\newcommand{\dimn}[1]{\mathrm{dim}(#1)}
\newcommand{\ip}[2]{\left\langle #1 \left| #2\right\rangle \right.}
\newcommand{\pov}[1]{\mathbf{#1}}
\newcommand{\ran}[1]{\mathrm{ran}(\mathbf{#1})}
\newcommand{\mat}[1]{\mathcal{M}_{#1}}
\newcommand{\inst}[3]{\mathcal{E}^{\mathbf{#1}}[#2](#3)}
\newcommand{\alg}[1]{\mathcal{#1}}
\newcommand{\abs}[1]{\lvert #1\rvert}
\newcommand{\ket}[1]{|#1\rangle}
\newcommand{\hil}[1]{\mathcal{#1}}
\newcommand{\co}[2]{\mathrm{co}_{\mathcal{#1}}(#2)^{-}}
\newcommand{\bh}{\mathcal{B}(\mathcal{H})}

\newcommand{\norm}[1]{\|#1\|}
\newcommand{\tr}[1]{\mathrm{Tr}(#1)}

\newtheorem{lemma}{Lemma}
\newtheorem*{lemma*}{Lemma}
\newtheorem{prop}[lemma]{Proposition}
\newtheorem{thm}{Theorem}
\newtheorem*{hjw}{HJW Theorem}
\theoremstyle{definition}
\newtheorem*{defn}{Definition}
\theoremstyle{remark}

\newtheorem*{notation}{Notation}

\title{Remote preparation of arbitrary ensembles and quantum bit commitment}
\author{Hans Halvorson \\ {\small Department of Philosophy,
  Princeton University} \\ {\small hhalvors@princeton.edu}}
\date{\today}
\begin{document}
\maketitle
\begin{abstract}
  The Hughston-Jozsa-Wootters theorem shows that any finite ensemble
  of quantum states can be prepared ``at a distance'', and it has been
  used to demonstrate the insecurity of all bit commitment protocols
  based on finite quantum systems without superselection rules.  In
  this paper, we prove a generalized HJW theorem for arbitrary
  ensembles of states on a $C^{*}$-algebra.  We then use this result
  to demonstrate the insecurity of bit commitment protocols based on
  infinite quantum systems, and quantum systems with Abelian
  superselection rules.
\end{abstract}

\section{Introduction}

Recent work in quantum cryptography has focused on questions of which
sorts of information-transfer protocols are secure from attempts at
cheating by an intruder or by one of the participants.  As early as
1984, the question was raised whether quantum theory would permit a
secure bit commitment protocol --- i.e., a protocol in which a bit of
information is committed by one party Alice to another party Bob, such
that Alice cannot change her commitment, and such that Bob cannot
determine Alice's commitment until given further information by Alice.
An initial protocol using pairs of polarized photons was proposed by
Bennett and Brassard \cite{bb84}; however, Bennett and Brassard showed
that this protocol can be cheated by exploiting the nonlocal
correlations of the EPR-Bohm state.

A number of other quantum bit commitment protocols have been proposed
in the intervening years (see \cite{brassard,bub} for reviews).  Most
of these protocols rely on the fact that a non-pure density operator
corresponds to more than one ensemble of quantum states.  In
particular, two different ensembles on a composite system can induce
the same density operator on a local system.  Thus, if Alice encodes
her bits into these two ensembles, then Bob cannot possibly determine
Alice's commitment until she provides further information about the
composite system.

However, Lo and Chau \cite{lo} and Mayers \cite{mayers1,mayers2} show
that, as a consequence of the Hughston-Jozsa-Wootters theorem, if a
bit commitment protocol is concealing against Bob, then it is not
binding against Alice.  (Kent's \cite{kent} relativistic bit
commitment protocol does not rely on the existence of alternative
decompositions of a density operator, and so its security is not
challenged by the Mayers-Lo-Chau result.)  That is, if the ensembles
prepared in the protocol are indistinguishable to Bob (i.e.,
correspond to approximately the same local density operator), then
Alice can ``steer'' between these ensembles after the Commit stage of
the protocol.
\begin{hjw}[\cite{hjw}]  Let $\hil{H}_{A}$ and $\hil{H}_{B}$ be finite-dimensional
  Hilbert spaces, let $\{ D_{i}\}_{i=1}^{n}$ be density operators on
  $\hil{H}_{B}$, and let $x$ be a unit vector in $\hil{H}_{A}\otimes
  \hil{H}_{B}$ such that $\mathrm{Tr}_{A}(P_{x})=\sum
  _{i=1}^{n}\lambda _{i}D_{i}$.  Then there are positive operators $\{
  A_{i}\}_{i=1}^{n}$ in $\mathcal{B}(\hil{H}_{A})\otimes I$ such that
\begin{equation} \ip{A_{i}^{1/2}x}{BA_{i}^{1/2}x}=
  \lambda _{i}\mathrm{Tr}(D_{i}B),\end{equation}
for all $B\in I\otimes \mathcal{B}(\hil{H}_{B})$.  \end{hjw}

Thus, the HJW theorem shows that any \emph{finite} decomposition of
$\mathrm{Tr}_{A}(P_{x})$ can be prepared from the state $P_{x}$ by a
measurement operation on $\hil{H}_{A}$.  However, all non-pure density
operators have countably infinite convex decompositions, as well as
uncountably infinite integral decompositions.  For the case of
countably infinite decompositions, Cassinelli et al.\ 
\cite{cassinelli} have proved a generalized HJW theorem; but their
results do not cover the case of integral decompositions.  What is
more, the HJW theorem and its generalization by Cassinelli et al.\ 
apply only to a very narrow class of quantum systems --- namely those
whose observables are represented by type I von Neumann factors.
Thus, these results do not directly establish the insecurity of bit
commitment protocols that employ systems with non-trivial
superselection rules (represented by direct sums of type I von Neumann
factors), or bit commitment protocols that employ infinite quantum
systems (represented by type II or type III von Neumann algebras).

In this paper, we prove a generalized HJW theorem for arbitrary
ensembles of states on a $C^{*}$-algebra.  We show first (in Section
II) that each measure on the state space of a $C^{*}$-algebra
$\alg{B}$ gives rise to a positive-operator valued measure with range
in the commutant $\alg{B}'$ of $\alg{B}$.  (This first result is
completely general, and does not impose any restrictions on the
$C^{*}$-algebra $\alg{B}$.)  We then show that when $\alg{B}'$ is a
hyperfinite von Neumann algebra, there is a completely positive
instrument that prepares the relevant ensemble of states on $\alg{B}$.
In Section III we apply our results to the question of the security of
bit commitment protocols.

\section{Generalized HJW theorem}

Our first result (Theorem \ref{tomita}) shows that for any
$C^{*}$-algebra $\alg{B}$ of operators acting on a Hilbert space
$\hil{H}$, a measure on the state space of $\alg{B}$ gives rise to a
corresponding POV measure with values in the commutant $\alg{B}'=\{
A\in \bh :[A,B]=0\;\mbox{for all}\;B\in \alg{B} \}$.  For the case
that $\alg{B}=I\otimes \mathcal{M}_{n}$, where $\mat{n}$ is the
$C^{*}$-algebra of $n\times n$ matrices over $\mathbb{C}$, our result
yields an alternate proof of the original HJW theorem.

Let $K$ denote the compact convex set of states of $\alg{B}$ with the
weak* topology.  [A net $\{ \omega _{a}\} _{a\in \mathbb{A}}$ of
states of $\alg{B}$ converges in the weak* topology to a state
$\omega$ just in case, for each $B\in \alg{B}$, $\lim _{a}\omega
_{a}(B)=\omega (B)$.  If $\alg{B}=\mat{n}$, then the weak* topology on
states is equivalent to the standard topology on density operators.]
In this paper, we consider positive regular measures on $(K,\Sigma )$,
where $\Sigma$ is the Borel $\sigma$-algebra of $K$.  When we say that
$\mu$ is a measure, it can be assumed that $\mu$ is positive and
regular.

\begin{defn}[\mbox{\cite[p.\ 12]{alfsen}}] If $\mu$ is a measure on
  the state space $K$, then the state
\begin{equation} \rho _{\mu}=\mu (K)^{-1}\int x\, d\mu (x)
  ,\end{equation}
is called the \emph{barycenter} of $\mu$. Measures $\mu$ and $\nu$ on
$K$ are said to be \emph{equivalent} if they have the same
barycenter.  \end{defn}

Let $K$ be the convex set of density operators on $\mathbb{C}^{n}$.
If $\rho$ is a density operator and $\mu$ is a finitely supported
measure on $K$ with barycenter $\rho$, then Hughston et al.\ call
$\mu$ a \mbox{$\rho$-\emph{ensemble}}.  So, the set of
$\rho$-ensembles consists of those measures on $K$ that have
barycenter $\rho$, and that are supported on a finite set.  In this
paper, we consider all measures with barycenter $\rho$, and not just
those with finite support.

\begin{notation} If $x$ is a vector in $\hil{H}$, we let $\omega _{x}(A)=\ip{x}{Ax}$,
  for all $A\in \bh$.  If $\alg{B}$ is a set of operators on
  $\hil{H}$, we let $\alg{B}x=\{ Bx:B\in \alg{B}\}$, and we let
  $[\alg{B}x]$ denote the closed linear span of $\alg{B}x$.
\end{notation}

\begin{lemma}[\mbox{\cite[Prop.~7.3.5]{kr}}] If $\alg{B}$ is a $C^{*}$-algebra of operators acting on
  the Hilbert space $\hil{H}$ and $\rho$ is a positive linear
  functional on $\alg{B}$ such that $\rho \leq \omega _{x}|_{\alg{B}}$
  for some vector $x$ in $\hil{H}$, then there is a positive operator
  $H$ in the unit ball of $\alg{B}'$ such that $\rho (A)=\omega
  _{x}(HA)=\omega _{H^{1/2}x}(A)$, for all $A\in \alg{B}$.
  \label{schwartz}
\end{lemma}

\begin{proof} Define a conjugate-bilinear functional $\varphi$ on
  $\alg{B}x$ by setting $\varphi (Ax,Bx)=\rho (A^{*}B)$.  Then,
\begin{equation} \abs{\varphi (Ax,Bx)}^{2}=\abs{\rho (A^{*}B)}^{2}\leq
  \rho (A^{*}A)\rho (B^{*}B)\leq
    \norm{Ax}^{2}\norm{Bx}^{2}.\end{equation}
The first inequality follows from the Cauchy-Schwartz inequality for
    the inner product $\langle A|B\rangle _{\rho}=\rho (A^{*}B)$ on
    $\alg{B}$.  Thus $\varphi$ is positive and bounded by $1$.  It follows that
    $\varphi$ has a unique extension to the subspace $[\alg{B}x]$.
    Moreover, the Riesz representation theorem entails that there is a
    positive operator $H$ on $[\alg{B}x]$ such that $\norm{H}\leq 1$
    and $\varphi (Ax,Bx)=\ip{Ax}{HBx}$.  In particular, $\rho
    (A)=\ip{x}{HAx}=\omega _{x}(HA)$ for all $A\in \alg{B}$.  Extend
    $H$ to the entire Hilbert space $\hil{H}$ by setting it to zero on
    $\hil{H}\ominus [\alg{B}x]$.  Since \begin{eqnarray}
\ip{Ax}{HCBx} &=&\rho (A^{*}CB)=\rho ((C^{*}A)^{*}B)\\
&=&\ip{C^{*}Ax}{HBx}=\ip{Ax}{CHBx} \end{eqnarray}
for all $C$ in $\alg{B}$, it follows that $[H,C]=0$ on $[\alg{B}x]$.
Since $[H,C]=0$ on $\hil{H}\ominus [\alg{B}x]$, it follows that
$[H,C]=0$ on the entire Hilbert space.  Therefore, $H\in \alg{B}'$.
\end{proof}

The following result is a special case of a theorem proved by Tomita
\cite{tomita} in 1956 (compare with \cite[Lemma 4.1.21,
Prop.~4.1.22]{br}).

\begin{thm} Let $\alg{B}$ be a $C^{*}$-algebra acting on the Hilbert
  space $\hil{H}$, and let $\mu$ be a probability measure on the state
  space of $\alg{B}$.  If there is a unit vector $x$ in $\hil{H}$ such
  that $\omega_{x}|_{\alg{B}}$ is the barycenter of $\mu$, then there
  is a POV measure $\pov{A}$ with range in $\alg{B}'$ such that
\begin{equation} \ip{A(S)^{1/2}x}{BA(S)^{1/2}x}=\int _{S}\omega (B)d\mu (\omega )
  ,\label{conditional} \end{equation} for all $S\in \Sigma$ and $B\in
  \alg{B}$. \label{tomita} \end{thm}

\begin{proof} 
  Let $S$ be a Borel subset of the state space of $\alg{B}$, and let
  $\rho _{S}=\int _{S}\omega d\mu (\omega )$.  Then $\rho _{S}$ is a
  positive linear functional on $\alg{B}$ with $\rho _{S}\leq \omega
  _{x}|_{\alg{B}}$.  By Lemma~\ref{schwartz}, there is a positive
  operator $A(S)$ in the unit ball of $\alg{B}'$ such that $\rho
  _{S}(B)=\omega _{x}(A(S)B)$ for all $B\in \alg{B}$, and $A(S)=0$ on
  $\hil{H}\ominus [\alg{B}x ]$.  In order to verify that $S\mapsto
  A(S)$ is countably additive, suppose that $\{ S_{i}:i\in \mathbb{N}
  \}$ are disjoint Borel subsets, and let $S=\cup
  _{i=1}^{\infty}S_{n}$.  Then for fixed $B\in \alg{B}$,
\begin{equation} \sum _{i=1}^{\infty}\chi _{S_{i}}(\omega )\cdot
  \omega (B) \:=\: \chi _{S}(\omega )\cdot \omega (B) ,\end{equation} 
and so the monotone convergence theorem entails that
\begin{equation}  \sum _{i=1}^{\infty} \left( \int _{S_{i}}\omega (B)\, d\mu (\omega
  ) \right) \:=\: \int _{S}\omega (B)\,d\mu (\omega )\:=\: \ip{x}{A(S)Bx}. \end{equation}
Furthermore, countable additivity of the map $Z\mapsto \ip{x}{ZBx}$ entails that
\begin{equation} \ip{x}{\left( \sum
  _{i=1}^{\infty}A(S_{i}) \right) Bx}\:=\: \sum _{i=1}^{\infty}\ip{x}{A(S_{i})Bx} .\end{equation}   
Replacing $B$ with $B^{*}C$, where $B,C\in \alg{B}$, we have
\begin{equation} \ip{Bx}{\sum _{i=1}^{\infty}A(S_{i})Cx} = \ip{Bx}{A(S)Cx},\end{equation}
and therefore $(\sum _{i=1}^{\infty}A(S_{i}))y=A(S)y$, for all $y\in
[\alg{B}x ]$.  Since $A(S)=0$ on $\hil{H}\ominus [\alg{B}x ]$, it
follows that $\sum _{i=1}^{\infty}A(S_{i})=A(S)$.
\end{proof}

Thus, if $\mu$ is a measure on the state space of Bob's algebra
$\alg{B}$, Alice's algebra $\alg{A}=\alg{B}'$ contains the range of a
POV measure $\pov{A}$ satisfying Eqn.\  \ref{conditional}.  But this
does not yet yield the conclusion that Alice can prepare the ensemble
$\mu$ on Bob's system --- for that, we need to show that Alice has an
``instrument'' corresponding to the POV measure $\pov{A}$.

\begin{defn}[\cite{davies,ozawa}] Let $(X,\Sigma )$ be a Borel space.
  A \emph{completely positive (CP) instrument} on $\bh$ is a map
  $\mathcal{E}:\Sigma \times \bh \rightarrow \bh$ such that:
\begin{enumerate}
\item for fixed $B\in \bh$, $\mathcal{E}[\; \cdot \; ](B)$ is
  $\sigma$-additive in the weak-operator topology;
\item for fixed $S\in \Sigma$, $\mathcal{E}[S](\; \cdot \;)$ is a
  completely positive linear map such that $\mathcal{E}[S](I)\leq I$.
\end{enumerate} 
\end{defn}

Since the map $\mathcal{E}[S](\;\cdot \;)$, $(S\in \Sigma )$, is
positive, it is automatically norm-continuous.  If, in addition, each
such map is weak-operator continuous on bounded sets, then $\alg{E}$
is said to be \emph{normal}.  [A net $\{ A_{a}\} _{a\in \mathbb{A}}$
of bounded operators on $\hil{H}$ converges in the weak-operator
topology to an operator $A$ just in case $\lim
_{a}\ip{x}{A_{a}y}=\ip{x}{Ay}$ for all vectors $x,y$ in $\hil{H}$.]
However, we do not require instruments to be normal, because
continuous PV measures give rise to non-normal instruments
\cite{davies,srinivas}, and a continuous ensemble of states on
$\alg{B}$ will give rise to a continuous PV measure in $\alg{B}'$.
  
Each instrument $\mathcal{E}$ determines a unique POV measure
$\pov{A}$ via the formula
\begin{equation} A(S)\equiv \inst{}{S}{I},\qquad
  (S\in \Sigma ). \label{compatible} \end{equation}  If Eqn.\  \ref{compatible} holds for an instrument $\mathcal{E}$ and a
  POV measure $\pov{A}$, then $\mathcal{E}$ and $\pov{A}$ are said to
  be \emph{compatible}.  For any given POV measure $\pov{A}$, there are many instruments that are
compatible with $\pov{A}$.  In fact, if $\Phi$ is a CP projection of $\bh$ with
$\mathrm{ran}(\Phi )\subseteq \ran{A}'$ then
\begin{equation} 
\inst{}{S}{B}=A(S)\Phi (B) ,\qquad (S\in \Sigma ,B\in \bh ),
\end{equation} is compatible with
$\mathbf{A}$. In particular, if $\rho$ is a state on $\bh$ then
\begin{equation}
\inst{}{S}{B}=A(S)\rho (B) ,\qquad (S\in \Sigma ,B\in \bh ),
\end{equation}
is compatible with $\pov{A}$.

Thus, given a POV measure $\pov{A}$ with range in Alice's algebra
$\alg{A}=\alg{B}'$, we can easily find an instrument $\mathcal{E}$
that is compatible with $\pov{A}$.  However, it does not follow that
Alice can in any sense measure $\pov{A}$ with the instrument
$\mathcal{E}$, because $\mathcal{E}$ may not be ``local'' to Alice's
system.  In particular, an instrument that is local to Alice's system
should not disturb the statistics of measurements of observables in
Bob's algebra $\alg{B}=\alg{A}'$.  In other words, for any state
$\rho$ on $\alg{B}$, the equation
\begin{equation} \rho \left( \inst{}{X}{B} \right) = \rho (B)
\label{local}  ,\end{equation}
should hold for all $B\in \alg{B}$.  But Eqn.\ \ref{local} holds for
all states $\rho$ on $\alg{B}$ iff the CP map $\inst{}{X}{\,\cdot \,}$
is the identity on $\alg{B}$.  Thus, we capture the locality
requirement with the following definition.

\begin{defn} An instrument $\alg{E}$ is \emph{local} to an
  algebra $\alg{A}$ just in case $\inst{}{S}{B}=\inst{}{S}{I}B$, for
  all $B\in \alg{A}'$ and $S\in \Sigma$.
\end{defn}

Of course, if $\pov{A}$ is a POV measure on $\mathbb{N}$, there is a
canonical instrument $\mathcal{E}^{\pov{A}}$ that is compatible with
$\pov{A}$ and local to any algebra containing $\ran{A}$:
\begin{equation} \inst{A}{S}{B}=\sum
  _{n\in S}A_{n}^{1/2}BA_{n}^{1/2} ,\qquad (S\subseteq \mathbb{N},B\in \bh ). \end{equation}
We wish to extend this result to show that for each POV measure $\pov{A}$ with
range in a $C^{*}$-algebra $\alg{A}$, there is a CP instrument
$\mathcal{E}^{\pov{A}}$ that is compatible with $\pov{A}$ and local to
$\alg{A}$.  In this paper, we prove this result for finite quantum
systems (Theorem~\ref{finite}), and for infinite quantum
systems that can be approximated, in an appropriate sense, by finite
quantum systems (Theorem~\ref{main}).  While our proof for the finite case
uses only elementary linear algebra, our proof for the infinite case
is non-constructive (i.e., invokes the axiom of choice in the form of
the Tychonoff product theorem), and uses tools from the theory of operator algebras.

We first show that if Alice has a finite quantum system, then she can
perform a ``maximally disturbing'' local operation --- i.e., an
operation that maps all her states to the maximally mixed state.

\begin{lemma} If $\alg{A}$ is finite-dimensional $C^{*}$-algebra on
  the Hilbert space $\hil{H}$ then there is a completely positive
  projection $\Phi$ from $\bh$ onto $\alg{A}'$.  In particular, $\Phi$
  maps $\alg{A}$ onto $\mathbb{C}I$.
  \label{project}
\end{lemma}

\begin{proof} (Compare with \cite[Prop.~8.3.11]{kr} and
  \cite{schwartz}.)  Since $\alg{A}$ is finite-dimensional,
  $\alg{A}=\bigoplus _{i=1}^{m}\alg{M}_{n(i)}$ for some positive
  integers $n(1),\dots ,n(m)$.  Consider the following statement:
\begin{quote} $(\dagger )$ There is a projective unitary representation
  $g\mapsto W(g)$ of a finite group $G$ in $\alg{A}$ such that $\{
  W(g):g\in G\}$ spans $\alg{A}$.
\end{quote} We first show that $(\dagger )$ holds when
$\alg{A}=\mat{n}$.  Let $\{ \ket{0},\dots ,\ket{n-1} \}$ be a basis
for $\mathbb{C}^{n}$, and for each $g\in \mathbb{Z}_{n}\times
\mathbb{Z}_{n}$ let $W(g)$ be the unitary operator on $\mathbb{C}^{n}$
defined by
\begin{equation}
W(g)\ket{a}=e^{ig_{1}a}\ket{a+g_{2}},\qquad (a=1,\dots ,n)
.\end{equation}
Then $g\mapsto W(g)$ is a projective representation of
$\mathbb{Z}_{n}\times \mathbb{Z}_{n}$ with
bicharacter $\xi (g,h)=e^{ig_{1}h_{2}}$; i.e., $W(g)W(h)=e^{ig_{1}h_{2}}W(g+h)$.  Furthermore, 
$\{ W(g):g\in \mathbb{Z}_{2}\times \mathbb{Z}_{2}\}$ is an orthonormal basis for
$\mat{n}$ relative to the inner product $\langle A|B\rangle
_{2}=\tr{A^{*}B}$.  Thus, we have established $(\dagger )$ for the case
  that $\alg{A}=\mat{n}$.  We now show that $(\dagger )$ holds when
  $\alg{A}=\bigoplus _{i=1}^{m}\mat{n(i)}$.  Indeed, let \begin{equation} G=\bigoplus
  _{i=1}^{m}\left[ \mathbb{Z}_{n(i)}\times \mathbb{Z}_{n(i)} \right]
  ,\end{equation} and take the direct sum of the corresponding projective
representations.
  
We now show that if $(\dagger )$ holds, then there is a CP projection
from $\bh$ onto $\alg{A}'$.  For each $g\in G$, define an automorphism
$\alpha _{g}$ of $\bh$ by
\begin{equation} \alpha _{g}(A)=W(g)^{*}AW(g),\qquad (B\in \bh
  ).\end{equation}
Then $\mathcal{G}=\{ \alpha _{g}:g\in G\}$ is a finite group of
automorphisms of $\bh$.  If $\alpha (A)=A$ for all $\alpha \in
\mathcal{G}$, then $AW(g)=W(g)A$ for all $g\in G$, and $A\in
\alg{A}'$.  Thus, 
\begin{equation}
\Phi (A)=\abs{\mathcal{G}}^{-1}\sum _{\alpha \in \mathcal{G}}\alpha (A) ,\qquad (A\in
\alg{B}(\hil{H})),\end{equation}
is a CP projection from $\bh$ onto $\alg{A}'$.  \end{proof}

We are now prepared to prove a generalized HJW theorem, valid for all
finite quantum systems (i.e., systems whose algebra of observables is
finite-dimensional).  In particular, if the $C^{*}$-algebra $\alg{B}'$
is finite-dimensional, then the product $\Phi \otimes \pov{A}$ of the
maximally disturbing operation $\Phi$ (from Lemma \ref{finite}) and
the POV measure $\pov{A}$ (from Theorem \ref{tomita}) yields an
instrument that prepares the ensemble $\mu$ on system $\alg{B}$.

\begin{thm}[Generalized HJW Theorem] Let $\alg{B}$ be a $C^{*}$-algebra acting on the Hilbert space
  $\hil{H}$, let $x$ be a unit vector in $\hil{H}$, and let $\mu$ be a
  measure on the state space of $\alg{B}$ such that $\omega
  _{x}|_{\alg{B}}$ is the barycenter of~$\mu$.  If $\alg{B}'$ is
  finite-dimensional then there is a CP instrument $\mathcal{E}$ on
  $\bh$ that is local to $\alg{B}'$ and
  \begin{equation} \ip{x}{\inst{}{S}{B}x}=\int _{S}\omega (B)d\mu
    (\omega ) , \end{equation}
\label{finite} for all $S\in \Sigma$ and $B\in \alg{B}$. 
\end{thm}

\begin{proof} By Lemma~\ref{finite}, there is a CP projection $\Phi$
  from $\bh$ onto $\alg{A}'$.  Let $\mathcal{E}=\Phi \otimes \pov{A}$,
  where $\pov{A}$ is the POV measure defined in Theorem~\ref{tomita}.
  That is,
\begin{equation} \inst{}{S}{B}=\Phi (B)A(S), \end{equation} for all $S\in \Sigma$ and $B\in \bh$.
\end{proof}
   
This generalized HJW theorem applies to bit commitment protocols that
employ continuous ensembles on finite quantum systems (e.g.,
continuous measures on the Bloch sphere), and to finite quantum
systems with Abelian superselection rules (direct sums of matrix
algebras).  However, this first result leaves open the possibility of
secure bit commitment protocols that employ infinite quantum systems.
So, in the following subsection, we prove a more general HJW theorem
that also applies to infinite quantum systems.

\subsection{HJW theorem for hyperfinite algebras}

\begin{defn} A von Neumann algebra $\alg{R}$ is said to be
  \emph{hyperfinite} just in case there is an upward directed family
  $\{\alg{R}_{a} \} _{a\in \mathbb{A}}$ of finite-dimensional
  $C^{*}$-algebras such that $\alg{R}$ is the weak-operator closure of
  $\bigcup _{a\in \mathbb{A}}\alg{R}_{a}$. \end{defn}

As in the finite case, an observer with a hyperfinite von Neumann
algebra can perform a maximally disturbing measurement operation.

\begin{lemma} If $\alg{R}$ is a hyperfinite von Neumann algebra acting
  on the Hilbert space $\hil{H}$ then there is a completely positive
  projection $\Phi$ from $\bh$ onto $\alg{R}'$. In particular, $\Phi$
  maps $\alg{R}$ onto $\mathbb{C}I$.  \label{humph}
\end{lemma}

\begin{notation} For an arbitrary operator $B$ in $\bh$ we write
  $\co{R}{B}$ for the weak-operator closed convex hull of $\{
  UBU^{*}:U\in \alg{R},\;U\:\text{unitary}\}$. \end{notation}

\begin{proof} (Compare with \cite[Prop.\ 8.3.11; Exercise 8.7.24]{kr} and
  \cite{schwartz}.)  Let $\{ \alg{R}_{a}:a\in \mathbb{A}\}$ be an
  increasing net of finite-dimensional $C^{*}$-algebras on $\hil{H}$
  such that
\begin{equation} \textstyle \left( \bigcup _{a\in \mathbb{A}}\alg{R}_{a}\right)
  ^{-}=\alg{R} ,\end{equation} where $\alg{X}^{-}$ denotes the
weak-operator closure of $\alg{X}$.  By Lemma~\ref{finite}, for each $a\in
\mathbb{A}$ there is a CP projection $\Phi _{a}$ from $\bh$
onto $\alg{R}_{a}'$ .  Let \begin{equation}
X=\prod _{B\in \bh }\co{R}{B} \end{equation} be the product
topological space, where each factor is equipped with the
weak-operator topology.  For fixed $B\in \bh$, $\co{R}{B}$ is weak-operator
closed and bounded, and is therefore weak-operator compact
\cite[Thm.\ 5.1.3]{kr}.  Thus, the Tychonoff product theorem
entails that $X$ is compact.  Let $\mathcal{M}$ be the subset of $X$ consisting
of mappings $\Phi$ that are positive, linear, normalized, and such
that
\begin{equation} \Phi (R_{1}'BR_{2}')=R_{1}'\Phi (B)R_{2}'
  ,\label{module} \end{equation}
for all $R_{1}',R_{2}'\in \alg{R}'$ and for all $B\in \bh$.  Since $\mathcal{M}$
is closed in $X$, $\mathcal{M}$ is compact, and $\{ \Phi _{a}:a\in
\mathbb{A}\}$ has a limit point $\Phi \in \mathcal{M}$.  Note that $\lim _{a}\Phi
_{a}=\Phi$ iff, for each fixed $B\in \bh$, $\wlim _{a}\Phi
_{a}(B)=\Phi (B)$.  We claim that $\Phi (B)\in
\alg{R}'$ for each $B\in \bh$.  Let $A\in \bigcup _{a\in \mathbb{A}}\alg{R}_{a}$; that is, there is an
  $m\in \mathbb{A}$ such that $A\in \alg{R}_{m}$.  Then, $A\Phi
  _{a}(B)=\Phi _{a}(B)A$, for all $a\geq m$.  Since the
  maps $Z\mapsto AZ$ and $Z\mapsto ZA$ are weak-operator continuous, 
\begin{eqnarray}
A\left[ \wlim _{a\geq m}\Phi _{a}(B) \right] &=& \wlim
_{a\geq m}\left[ A\Phi _{a}(B)\right] \:=\:\wlim _{a\geq m}\left[ \Phi
  _{a}(B)A\right] \\
&=& \left[ \wlim _{a\geq m}\Phi _{a}(B)\right] A
.\end{eqnarray} Since $\Phi (B)=\wlim _{a}\Phi _{a}(B)=\wlim _{a\geq m}
\Phi _{a}(B)$, it follows that $A\Phi (B)=\Phi (B)A$.  Since $A$ was an arbitrary element
of $\bigcup _{a\in \mathbb{A}}\alg{R}_{a}$, $\Phi (B)\in ( \bigcup _{a\in
  \mathbb{A}}\alg{R}_{a})'=\alg{R}'$.  Therefore $\mathrm{ran}(\Phi
)=\alg{R}'$.  Finally, since $\Phi$ is an $\alg{R}'$-bimodule mapping (i.e.,
Eqn.\  \ref{module} holds), $\Phi$ is idempotent and completely positive \cite[Cor.~3.4]{takesaki}.  \end{proof}

Again, a maximally disturbing operation $\Phi$ can be tensored with
the POV measure $\pov{A}$ to yield an instrument that prepares the
ensemble $\mu$ on Bob's system.

\begin{thm}[Generalized HJW Theorem] Let $\alg{B}$ be a $C^{*}$-algebra acting on the Hilbert space
  $\hil{H}$, let $x$ be a unit vector in $\hil{H}$, and let $\mu$ be a
  measure on the state space of $\alg{B}$ such that $\omega
  _{x}|_{\alg{B}}$ is the barycenter of~$\mu$.  If $\alg{B}'$ is
  hyperfinite then there is a CP instrument $\mathcal{E}$ on $\bh$
  that is local to $\alg{B}'$ and
  \begin{equation} \ip{x}{\inst{}{S}{B}x}=\int _{S}\omega (B)d\mu
    (\omega ) , \end{equation}
\label{main} for all $S\in \Sigma$ and $B\in \alg{B}$. 
\end{thm}

\begin{proof} The proof is identical to the
  proof of Theorem~\ref{finite}, with Lemma~\ref{humph} replacing
  Lemma~\ref{project}. \end{proof}

\section{Application to bit commitment}

The Mayers-Lo-Chau theorem shows that when $\alg{A}=\mat{n}\otimes I$
and $\alg{B}=\alg{A}'$, and when bits are encoded in finite ensembles,
then $(\alg{A},\alg{B})$ cannot be used to implement a secure bit
commitment protocol.  The generalized HJW theorem allows us to extend
this result to the case where $\alg{A}$ $(=\alg{B}')$ is an arbitrary
hyperfinite von Neumann algebra, and to encodings that employ
arbitrary ensembles of states on $\alg{B}$.  In particular, the
generalized HJW theorem entails that there can be no secure bit
commitment protocol using infinite (hyperfinite) quantum systems, or
quantum systems with Abelian superselection rules.

\subsection{Bit commitment with infinite quantum systems}

The quantum bit commitment protocols that have been proposed to date
employ finite quantum systems.  In this subsection, we describe a bit
commitment protocol that employs continuous ensembles of states on
infinite qubit lattices.  Since this protocol does not fall within the
range of validity of the HJW theorem, it is immune to current no-go
theorems against bit commitment.  However, we show that this protocol
can be cheated by exploiting the non-local correlations of an
``infinitely entangled'' EPR state (see \cite{werner}).

Let $\ket{0,0}$ and $\ket{0,1}$ be orthogonal unit eigenvectors of
$\sigma _{x}$, and let $\ket{1,0}$ and $\ket{1,1}$ be orthogonal unit
eigenvectors of $\sigma _{y}$.  Then, heuristically, the states of a
one-dimensional infinite qubit lattice include vectors of the form
\begin{equation} \latt{b}{s}\: =_{\mathrm{def}}\: \otimes _{i=1}^{\infty} \ket{b,s(i)}
  ,\qquad (s\in (\mathbb{Z}_{2})^{\omega}) ,\end{equation} with
$b=0$ or $b=1$. (We provide a rigorous definition of these states below.)

During the Commit stage of the protocol, Alice performs operations on
a composite $(\alg{A},\alg{B})$ consisting of two lattice systems
$\alg{A}$ and $\alg{B}$, and she then sends system $\alg{B}$ to Bob.
During the Unveil stage, Alice makes measurements on $\alg{A}$, and
sends classical information to Bob, who then makes measurements on
$\alg{B}$.

\bigskip \noindent
\fbox{\parbox[c]{0.97\textwidth}{\begin{description}
    \item[Commit:] For $b=0,1$, Alice chooses a random sequence $s\in
      (\mathbb{Z}_{2})^{\omega}$, and prepares the state
\[ \latt{b}{s} _{A}
\otimes \latt{b}{s} _{B} .\] Alice holds part $A$, and sends part $B$
to Bob.  (So, the ensemble Bob receives is an equal mixture over
$\latt{b}{s}$, for $s\in (\mathbb{Z}_{2})^{\omega}$.)
\item[Unveil:] Alice measures the observable
\[ A_{b} = \sum _{i=1}^{\infty}\frac{2}{3^{i}}P_{b}^{(i)}
, \] on her systems, where $P_{b}=\frac{1}{2}(I+\sigma _{b})$ and
\[ P_{b}^{(i)}=I\otimes \cdots I\otimes P_{b}\otimes
I\otimes \cdots .\] (Each state $\latt{b}{s}$ is an eigenstate of
$A_{b}$, and when $s_{1}\neq s_{2}$, $\latt{b}{s_{1}}$ and
$\latt{b}{s_{2}}$ assign different values to $A_{b}$.)  Alice sends
the results of her measurements (a list of numbers in the Cantor set)
to Bob.  Bob measures $A_{b}$ on his systems and compares his numbers
with Alice's.  Bob accepts if the two lists agree, and rejects if the
two lists disagree.
\end{description} } }

\bigskip \noindent Let $\rho _{b}$ be the state that Bob receives.  It
is intuitively clear that if Alice follows the protocol honestly then
$\rho _{0}=\rho _{1}$, and so Bob cannot cheat.  (We prove this fact
below.)

We now tighten up the mathematical description of the systems involved
in the protocol.  The observables of a one-dimensional qubit lattice
are represented by the $C^{*}$-algebraic infinite direct product
\begin{equation} \alg{A}=\bigotimes _{i\in \mathbb{N}}\mat{n(i)}  ,\end{equation}
where $n(i)=2$ for each $i\in \mathbb{N}$.  For each $i\in \mathbb{N}$
and $A\in \mat{2}$, let
\begin{equation} A^{(i)}=I\otimes \cdots \otimes I \otimes A \otimes I
  \cdots ,\end{equation}
where $A$ is in the $i$-th position.  If for each $i\in \mathbb{N}$,
$\omega _{i}$ is a state of $\mat{2}$, then there is a unique state $\otimes _{i=1}^{\infty}\omega _{i}$ of
$\alg{A}$ defined by
\begin{equation}
\left( \otimes _{i=1}^{\infty}\omega _{i} \right) \left( A^{(j)}
\right) = \omega _{j}(A).\end{equation}  Furthermore, $\otimes
_{i=1}^{\infty}\omega _{i}$ is pure iff each $\omega _{i}$ is pure, and is a
trace iff each $\omega _{i}$ is a trace \cite[Prop.~11.4.7]{kr}.  Thus, if $\{ \ket{i}:i\in
\mathbb{N}\}$ are unit vectors in $\mathbb{C}^{2}$, then $\otimes
_{i=1}^{\infty}\ket{i}$ can be used to denote the corresponding
pure state of $\alg{A}$.  In particular, for any $s\in
(\mathbb{Z}_{2})^{\omega}$, $\latt{b}{s}$ does in fact correspond to a pure state of $\alg{A}$.     

Let $\alg{B}$ be an isomorphic copy of $\alg{A}$.  Since $\alg{A}$ is
a uniform limit of an increasing sequence of finite-dimensional
algebras, it is nuclear; i.e., there is a unique norm on the algebraic
tensor product $\alg{A}\odot \alg{B}$ whose completion is a
$C^{*}$-algebra.  We denote this $C^{*}$-algebra by $\alg{A}\otimes
\alg{B}$.  We now establish the existence of the ensembles described
in the protocol, and we show that they give rise to the same quantum
state (namely, the ``maximally mixed'' tracial state) on system
$\alg{B}$.

\begin{prop}  If $\mu$ is the normalized Haar measure on
  $(\mathbb{Z}_{2})^{\omega}$ then there is a probability measure $\mu
  _{b}$ on the state space of $\alg{A}\otimes \alg{B}$ such that
  \begin{equation} \mu _{b}\left( \left\{ \latt{b}{s} _{A}\otimes
        \latt{b}{s} _{B}:s\in S \right\} \right) = \mu (S)
    ,\label{agree} \end{equation} for every Borel subset $S$ of $\cantor$.  Furthermore, if $\rho _{b}$ is the
  barycenter of $\mu _{b}$ then $\rho _{b}|_{I\otimes \alg{B}}$ is the
  tracial state. \label{spins}
\end{prop}
  
To establish the first part of Proposition~\ref{spins}, it will
suffice to show that
\begin{equation} s\;\stackrel{\varphi}{\longmapsto} \;\latt{b}{s} _{A}\otimes \latt{b}{s} _{B} ,\end{equation}
is a continuous mapping of $\cantor$ into the state space of
$\alg{A}\otimes \alg{B}$ (with the weak* topology).  For then the
induced measure $\mu _{b}=\mu \circ \varphi ^{-1}$ will satisfy Eqn.\ 
\ref{agree}.

Let $G=\sum _{i\in \mathbb{N}}(\mathbb{Z}_{2}\oplus
\mathbb{Z}_{2})_{i}$ be the direct sum of a countable number of copies
of $\mathbb{Z}_{2}\oplus \mathbb{Z}_{2}$.  Elements of $G$ are
sequences with values in $\mathbb{Z}_{2}\oplus \mathbb{Z}_{2}$ that
differ from the identity $(0,0)$ in only finitely many positions.  Let
$V_{(0,0)}=I$, $V_{(0,1)}=\sigma _{x}$, $V_{(1,0)}=\sigma _{y}$, and
$V_{(1,1)}=\sigma _{z}$, and for any $s\in G$, let
\begin{equation}
U(s)\:=_{\mathrm{def}}\:V_{s(1)}\otimes V_{s(2)}\otimes V_{s(3)}\otimes \cdots \in \alg{A}
.\end{equation}
Then the set $\{ U(s):s\in G\}$ is linearly dense in $\alg{A}$.  
Let $H_{b}$ denote the subgroup of $G$ generated by
those sequences $s$ with the property that $s(i)=(0,0)$ or
$s(i)=(b,b\oplus 1)$ for all $i\in \mathbb{N}$.  Then, $\{ U(s):s\in
H_{b}\}$ generates an Abelian subalgebra of $\alg{A}$; namely, the
algebra generated by the spin operator $V_{(b,b\oplus 1)}$ at each lattice site.

\begin{lemma} If $\alg{C}$ is the Abelian subalgebra of $\alg{A}$ generated by
  $\{ U(s):s\in H_{b} \}$ then the pure state space of $\alg{C}$ is
  homeomorphic to $(\mathbb{Z}_{2})^{\omega}$.
\end{lemma}

\begin{proof} The Abelian $C^{*}$-algebra $\alg{C}$ is
  isomorphic to the $C^{*}$-algebra $C(X)$ of continuous
  complex-valued functions on $X$, where $X$ is the pure state space
  of $\alg{C}$ equipped with the weak* topology.  Furthermore, if
  $C(X)$ and $C(Y)$ are isomorphic then $X$ and $Y$ are homeomorphic.
  Thus, if $\alg{C}\simeq C(Y)$ then the space of pure states of
  $\alg{C}$ is homeomorphic to $Y$.  Now, $\alg{C}\simeq \bigotimes
  _{i=1}^{\infty}\alg{N}_{i}$, where $\alg{N}_{i}$ is the Abelian
  algebra generated by $\sigma _{b}$.  Since $\alg{N}_{i}$ is
  isomorphic to $C(\mathbb{Z}_{2})$,
\begin{equation} \bigotimes _{i=1}^{\infty}\alg{N}_{i}\:\simeq
  \:\bigotimes _{i=1}^{\infty}C(\mathbb{Z}_{2})\:\simeq \:C(\cantor )
  ,\end{equation} where
  $\cantor$ is equipped with the product topology (see
  \cite[pp.\ 910--911; Prop.\ 11.4.3]{kr}).  Therefore
  the pure state space of $\alg{C}$ is homeomorphic to $\cantor$. 
\end{proof} 
  
Since $\alg{C}$ is isomorphic to $C(\cantor )$, there is (by the Riesz
representation theorem) a one-to-one correspondence between positive
normalized measures on $\cantor$ and states on $\alg{C}$.

\begin{lemma} If $\mu$ is the Haar measure on $\cantor$ then the
  barycenter of $\mu$ is $\tau |_{\alg{C}}$, where $\tau$ is the trace
  on $\alg{A}$. \end{lemma}

\begin{proof} Let $\sigma (\alg{C})$ denote the pure state space of $\alg{C}$,
  and let $\rho =\int _{\sigma (\alg{C})}\omega d\mu (\omega )$ be the
  barycenter of $\mu$.  To show that $\rho =\tau$, it will suffice to
  show that $\rho (U(s))=0$ whenever $s\in H_{b}-\{ e\}$.  Indeed, if
  $s\neq e$ then there is an $i\in \mathbb{N}$ such that
  $s(i)=(b,b\oplus 1)$.  Let $s'$ be the element of $\oplus
  _{i=1}^{\infty}(\mathbb{Z}_{2}\oplus \mathbb{Z}_{2})_{i}$ such that
  $s'(j)=s(j)$ when $j\neq i$, and $s'(i)=(b\oplus 1,b)$.  Then
  $U(s')^{*}U(s)U(s')=-U(s)$.  Since $\mu$ is translation-invariant,
  $\rho (U(s))=-\rho (U(s))$.  Therefore $\rho (U(s))=0$.  \end{proof}
  
\begin{lemma} There is a completely positive projection $\Phi$ from
  $\alg{A}$ onto $\alg{C}$ such that $\tau (\Phi(A))=\tau (A)$ for all
  $A$ in $\alg{A}$.  \label{measure} \end{lemma}

\begin{proof} For each $t\in \mathbb{R}$, define an automorphism $\alpha _{t}$ of
  $\alg{A}$ by
\begin{equation}
\alpha _{t}(B)=e^{-itA_{b}}Be^{itA_{b}} ,\qquad (B\in \alg{A})
.\label{gert} \end{equation} Since $A_{b}$ is bounded, the map $t\mapsto \alpha
_{t}(B)$ is norm-continuous.  If $\nu$ is an
invariant mean on $\mathbb{R}$, then 
\begin{equation} \Phi(B)=\int _{\mathbb{R}}\alpha _{t}(B)\, d\nu (t)
  ,\qquad (B\in \alg{A}),\label{collapse} \end{equation}
is a positive linear map on $\alg{A}$ \cite[Lemma
7.4.4]{petersen}.  (To show that $\Phi$ is completely positive, it
will suffice to show that the range of $\Phi$ is Abelian.)  Clearly $\Phi(C_{1}BC_{2})=C_{1}\Phi(B)C_{2}$ for all $C_{1},C_{2}\in
\alg{C}$, and $B\in \alg{A}$.  In particular, $\Phi(C)=C$ for
all $C\in \alg{C}$.  To see that the image of $\Phi$ lies in $\alg{C}$, let $s$ be an
element of \mbox{$G=\sum _{i\in \mathbb{N}}(\mathbb{Z}_{2}\oplus
  \mathbb{Z}_{2})_{i}$}.  If $s\in H_{b}$ then $U(s)\in \alg{C}$ and
$\Phi (U(s))=U(s)$.  Suppose then that $s\not\in H_{b}$; that is,
there is an $i\in \mathbb{N}$ such that either $s(i)=(b\oplus 1,b)$ or
$s(i)=(1,1)$.  (It will suffice to consider the first case; the second
case follows by symmetry.)  Then
\begin{equation}
U(s)= V_{s(1)}\otimes \cdots \otimes V_{s(i-1)}\otimes V_{s(i)}\otimes V_{s(i+1)}
\otimes \cdots ,\end{equation}
and 
\begin{equation} 
\Phi (U(s)) = V_{s(1)}\otimes \cdots \otimes V_{s(i-1)}\otimes B_{i}\otimes V_{s(i+1)}
\otimes \cdots ,\end{equation}
where 
\begin{equation}
B_{i}= \int _{\mathbb{R}} e^{-itP_{b}}
  \left[ V_{(b\oplus 1,b)} \right] e^{itP_{b}} d\nu (t) = 0.
\end{equation}
Thus, $\Phi (U(s))=0$.  Since $\{ U(s):s\in G\}$ spans $\alg{A}$, it
follows that $\mathrm{ran}(\Phi )=\alg{C}$.  To see that $\tau
=\tau\circ \Phi$, note that every non-identity element of $\{
U(s):s\in G\}$ is trace-free.  If $s\in H_{b}$, then $\Phi
(U(s))=U(s)$ and therefore $\tau (\Phi (U(s)))=\tau (U(s))$.  If
$s\not\in H_{b}$, then $\Phi (U(s))=0$ and $\tau (U(s))=0$.  Since
$\tau$ and $\tau \circ \Phi$ are continuous linear functionals, $\tau
=\tau\circ \Phi$.
\end{proof}
  
The mapping $\Psi =\Phi \otimes \Phi$ is a CP projection from
$\alg{A}\otimes \alg{B}$ onto $\alg{C}\otimes \alg{C}$, and its
adjoint $\Psi ^{*}$ is a weak* continuous mapping from the state space
of $\alg{C}\otimes \alg{C}$ into the state space of $\alg{A}\otimes
\alg{B}$.  Let $\sigma (\alg{C}\otimes \alg{C})$ denote the pure state
space of $\alg{C}\otimes \alg{C}$, and let $\sigma (\alg{A}\otimes
\alg{B})$ denote the pure state space of $\alg{A}\otimes \alg{B}$.
Using $\Psi ^{*}$ again to denote the restriction of $\Psi ^{*}$ to
$\sigma (\alg{C}\otimes \alg{C})$, and identifying $\ps{\alg{C}\otimes
  \alg{C}}$ with $\cantor \times \cantor$, it follows that $\Psi ^{*}$
is a continuous injection of $\cantor \times \cantor$ into $\sigma
(\alg{A}\otimes \alg{B})$.  Note that
\begin{equation} \Psi ^{*}[(s,s)]= \latt{b}{s}_{A}\otimes \latt{b}{s} _{B}\end{equation} 
and so the mapping 
\begin{equation} s\stackrel{\varphi}{\longmapsto} \latt{b}{s}_{A}\otimes \latt{b}{s}_{B}=(\Psi ^{*}\circ \Delta )(s) ,\end{equation}
where $\Delta (s)=(s,s)$, is continuous, which establishes the first
part of Proposition~\ref{spins}.

Now let $\rho _{b}$ denote the barycenter of $\mu _{b}$, and let $\nu
_{b}=_{\mathrm{def}}\mu \circ (\Phi ^{*})^{-1}$ denote the measure on
$\sigma (\alg{B})$ induced by $\Phi ^{*}$ from the measure $\mu$ on
$\sigma (\mathcal{C})$.  Then for any $B\in \alg{B}$,
\begin{eqnarray} 
\rho _{b}(I\otimes B)&=& \int _{\sigma (\alg{A}\otimes \alg{B})}\omega
(I\otimes B)d\mu _{b}(\omega ) \:=\: \int _{\sigma (\alg{B})} \omega (B)d\nu _{b} (\omega ) \\
&=& \int _{\sigma (\alg{C})}\omega (\Phi (B))d\mu (\omega ) \:=\: \tau
(\Phi (B)) \:=\: \tau (B). \end{eqnarray}
This establishes the second part of Proposition~\ref{spins}.  Thus, $\mu _{0}$ and $\mu _{1}$ are
the ensembles prepared by Alice if she
follows the protocol honestly. 

Finally, we show that Alice can cheat by preparing an entangled state
during the Commit stage rather than $\mu _{0}$ or $\mu _{1}$.  In
particular, if for each $i\in \mathbb{N}$, $\psi _{i}=\psi$ is the
Bohm-EPR state of $\mat{2}\otimes \mat{2}$, then $\omega
=_{\mathrm{def}}\otimes _{i=1}^{\infty}\psi _{i}$ is a pure state of
$\bigotimes _{i=1}^{\infty}\left( \mat{i}\otimes \mat{i} \right) =
\alg{A}\otimes \alg{B}$ \cite{werner}.  It is not difficult to see,
then, that if Alice performs a nonselective measurement of $A_{b}$
(represented by the CP map in Eqn.\ \ref{gert}) when $\alg{A}\otimes
\alg{B}$ is in state $\omega$, then the posterior state is the
ensemble $\mu _{b}$.  Therefore, if Alice prepares $\omega$ during the
Commit stage, then she can unveil either $0$ or $1$.

\subsection{Bit commitment and superselection rules}

It has recently been argued by Mayers, Kitaev, and Preskill
\cite{kitaev,mayers3}, in response to a question raised by Popescu
\cite{popescu}, that the no-go theorem for bit commitment extends to
the case of quantum systems with superselection rules.  The
generalized HJW theorem provides another route to this result, at
least for systems whose superselection rules are Abelian.  In the case
of Abelian superselection rules, $\alg{A}=\alg{B}'$; that is, Alice
can perform any operation that commutes with Bob's measurement
operations.  And the generalized HJW theorem shows that an observer
with algebra $\alg{B}'$ can steer system $\alg{B}$ into any ensemble
consistent with $\omega _{x}|_{\alg{B}}$.  Thus, a bit commitment
protocol is perfectly concealing against Bob only if it is not binding
against Alice.  However, the generalized HJW theorem has nothing to
say (directly) about Alice's ability to cheat when both systems are
governed by non-Abelian superselection rules (in which case
$\alg{A}\subset \alg{B}'$).  

Mayers et al.\ \cite{kitaev} claim that --- HJW theorem aside ---
Alice can always steer Bob's system into the state of her choice by
adding, if necessary, an appropriate ancilla to her system.  Their
argument is based on a more general claim that restrictions imposed by
superselection rules on a local system can always be effectively
removed by embedding the local system in a larger system (in
particular, by adding an ancilla).

The formalism of elementary quantum mechanics imposes no restriction
on adding ancillae.  However, in the setting of algebraic quantum
field theory, an observer can measure only those observables that
correspond to her spacetime region.  As a result, adding ancillae is
not permitted --- at least if ``adding an ancilla'' is interpreted to
mean that Alice can measure observables that are not in her local
observable algebra $\alg{R}(O_{A})$.  Thus, in this richer theoretical
framework, Alice is subject to further constraints on her ability to
simulate any operation that commutes with Bob's measurement
operations, and these constraints could --- it seems theoretically
possible --- prevent Alice from cheating in a bit commitment protocol.
(It would be interesting to explore connections between the formal
condition $\alg{A}\subset \alg{B}'$ and relativistic constraints of
the sort exploited by Kent's \cite{kent} bit commitment protocol.)

\subsection{Limitations on the generalized HJW theorem}

Let us say that a bit commitment protocol employs a \emph{quantum
  encoding} just in case Alice encodes her choice of a bit $0$ or $1$
in two ensembles $\mu _{0}$ or $\mu _{1}$ of quantum states.  Then,
even in the case of bit commitment schemes that employ quantum
encodings, there is one further assumption of the generalized HJW
theorem that is not \textit{prima facie} guaranteed to hold in any bit
commitment protocol: the assumption that the barycenter of $\mu _{b}$
is a \emph{vector} state.  (Let us call this latter assumption the
\textit{vector state assumption}.)

First, it is not difficult to find pairs of $C^{*}$-algebras
$(\alg{A},\alg{B})$, and measures $\mu _{b}$ on the state space of
$\alg{B}$ such that the vector state assumption does not hold: e.g.,
let $\alg{B}=\mat{2}$, and let $\mu _{b}$ be the measure on the state
space of $\alg{B}$ that assigns $\frac{1}{2}$ to each of
$\frac{1}{2}(I+\sigma _{b})$ and $\frac{1}{2}(I-\sigma _{b})$.  (Of
course, this trivial example could not be used to construct a secure
bit commitment protocol, since Alice could not perform any non-trivial
measurements to verify her commitment to Bob.)  However, the vector
state assumption does hold when $\alg{B}$ has a separating vector in
$\hil{H}$.

\begin{defn} A vector $x$ in the Hilbert space $\hil{H}$ is said to be
  \emph{separating} for the $C^{*}$-algebra $\alg{B}$ just in case
  $Bx=0$ only if $B=0$ for all $B\in \alg{B}$.  \end{defn}

\begin{prop}[\mbox{\cite[Thm.\ 7.3.8]{kr}}] If $\alg{B}$ is a
  $C^{*}$-algebra acting on the Hilbert space $\hil{H}$ and if
  $\alg{B}$ has a separating vector $x$ in $\hil{H}$, then each state
  of $\alg{B}$ is implemented by some vector in $\hil{H}$.
\end{prop}

Thus, if $\alg{B}$ has a separating vector in the Hilbert space
$\hil{H}$ (and if $\alg{A}=\alg{B}'$) then any ensemble of states on
$\alg{B}$ corresponds to a state $\omega _{x}|_{\alg{B}}$ induced by a
vector $x$ in $\hil{H}$, and the generalized HJW theorem entails that
any two equivalent ensembles can be prepared at a distance (from a
common state).  For example, $\alg{B}=I_{A}\otimes
\alg{B}(\hil{H}_{B})$ has a separating vector in $\hil{H}_{A}\otimes
\hil{H}_{B}$ if and only if $\dimn{\hil{H}_{B}}\leq
\dimn{\hil{H}_{A}}$ \cite{superentangled}.  So, in the case of
elementary quantum systems, by adding an ancilla, Alice can ``make her
Hilbert space as large as Bob's'', which ensures that their joint
Hilbert space $\hil{H}=(\hil{H}_{A'}\otimes \hil{H}_{A})\otimes
\hil{H}_{B}$ contains a vector representative of each of Bob's states.

Nonetheless, there are $C^{*}$-algebras that do not have --- and could
not have, in any faithful representation --- a separating vector,
e.g., $C^{*}$-algebras which contain an uncountable family of mutually
orthogonal projection operators.  But if $\alg{B}$ does not have a
separating vector, then the HJW theorem doesn't show that an observer
with algebra $\alg{B}'$ could perform operations that prepare any one
of two equivalent measures on the state space of $\alg{B}$ (from a
common ancestor state).  Until the HJW theorem is generalized to cover
such cases, there remains a small, but theoretically crucial, loophole
in current proofs of the impossibility of secure bit commitment.

\section{Conclusion}

We have shown, subject to a mild constraint (viz., that the systems
involved are ``hyperfinite''), that any two equivalent measures on the
state space of a $C^{*}$-algebra can be prepared ``at a distance''.
This result generalizes the Hughston-Jozsa-Wootters theorem, and so
can be used to extend the Mayers-Lo-Chau argument against the security
of quantum bit commitment protocols.

However, the results proved to date --- including the results in this
paper --- are not yet sufficient to rule out the security of any
conceivable quantum bit commitment protocol.  First, it remains an
open question whether an analogue of the HJW theorem holds for any
system whose observables can be represented by self-adjoint operators
in some abstract (not necessarily nuclear) $C^{*}$-algebra.  Second,
in order to invoke the HJW theorem in an argument against bit
commitment, one must make further physical assumptions --- e.g., that
the states on Bob's system correspond to vector states of some larger
system $S$, and that Alice can perform any operation on $S$ that
commutes with Bob's measurement operations --- that have yet to be
justified in a fully general context.

\begin{center} {\bf Acknowledgments } \end{center} 

This work was motivated by discussions with Jeff Bub and the late Rob
Clifton.  Thanks to Jens Eisert and Hans Maassen for helpful
correspondence, and to Mary Beth Ruskai and an anonymous referee for
comments on an earlier draft.


\begin{thebibliography}{99} 
  
\bibitem{alfsen} Alfsen, E., {\it Compact Convex Sets and Boundary
    Integrals} (Springer, New York, 1971).
  
\bibitem{bb84} Bennett, C. and Brassard, G., ``Quantum cryptography:
  Public key distribution and coin tossing, '' in {\it Proceedings of
    IEEE International Conference on Computers, Systems, and Signal
    Processing} (IEEE, 1984), pp. 175--179.
  
\bibitem{bcjl} Bennett, C., Cr{\'e}peau, C., Jozsa, R. and Langlois,
  D., ``A quantum bit commitment scheme provably unbreakable by both
  parties,'' {\it Proceedings of the 34th Annual IEEE Symposium on the
    Foundations of Computer Science.} 1993, pp. 362--371.
  
\bibitem{brassard} Brassard, G., Cr{\'e}peau, C., Mayers, D., and
  Salvail, L., ``A brief review on the impossibility of quantum bit
  commitment,'' quant-ph/9712023
  
\bibitem{br} Bratteli, O. and Robinson, D., {\it Operator Algebras and
    Quantum Statistical Mechanics}, Vol. 1 (Springer, New York, 1987).
  
\bibitem{bub} Bub, J., ``The quantum bit commitment theorem,'' Found.\ 
  Phys.\ {\bf 31} 735--756 (2001).
  
\bibitem{cassinelli} Cassinelli, G., De Vito, E. and Levrero, A., ``On
  the decompositions of a quantum state,'' J. Math.\ Anal.\ Appl.\ 
  {\bf 210}, 472--483 (1997).
  
\bibitem{superentangled} Clifton, R., Feldman, D., Halvorson, H.,
  Redhead, M. and Wilce, A., ``Superentangled states,'' Phys.\ Rev.\ A
  {\bf 58}, 135--145 (1998).
  
\bibitem{davies} Davies, E., {\it Quantum Theory of Open Systems}
  (Academic, New York, 1976).
  
\bibitem{hjw} Hughston, L., Jozsa, R. and Wootters, W., ``A complete
  classification of quantum ensembles having a given density matrix,''
  Phys.\ Lett.\ A {\bf 183}, 14--18 (1993).
  
\bibitem{kent} Kent, A., ``Unconditionally secure bit commitment,''
  Phys.\ Rev.\ Lett.\ {\bf 83}, 1447--1450 (1999).
  
\bibitem{werner} Keyl, M., Schlingemann, D. and Werner, R.,
  ``Infinitely entangled states.'' Quantum Inf.\ Comput.\ {\bf 3},
  281--306 (2003).
  
\bibitem{kr} Kadison, R. and Ringrose, J., {\it Fundamentals of the
    Theory of Operator Algebras}.  (American Mathematical Society,
  Providence, RI, 1997).
  
\bibitem{kitaev} Kitaev, A., Mayers, D. and Preskill, J.,
  ``Superselection rules and quantum protocols,'' quant-ph/0310088
  
\bibitem{lo} Lo, H.-K. and Chau, H. F., ``Is quantum bit commitment
  really possible?'' Phys.\ Rev.\ Lett.\ {\bf 78}, 3410--3413 (1997).
  
\bibitem{mayers1} Mayers, D., ``Unconditionally secure quantum bit
  commitment is impossible,'' in \textit{Proceedings of the Fourth
    Workshop on Physics and Computation} (Boston, 1996), pp. 224--228.
  
\bibitem{mayers2} Mayers, D., ``Unconditionally secure quantum bit
  commitment is impossible,'' Phys.\ Rev.\ Lett.\ {\bf 78}, 3414--3417
  (1997).
  
\bibitem{mayers3} Mayers, D., ``Superselection rules in quantum
  cryptography,'' quant-ph/0212159
  
\bibitem{ozawa} Ozawa, M., ``Quantum measuring processes of continuous
  observables,'' J. Math. Phys. {\bf 25}, 79--87 (1984).
  
\bibitem{petersen} Petersen, G., {\it $C^{*}$-algebras and Their
    Automorphism Groups}. (Academic Press, NY, 1979).
  
\bibitem{popescu} Popescu, S., ``Multi-party entanglement,'' MSRI
  lecture, Dec 2002.
  
\bibitem{schwartz} Schwartz, J., ``Two finite, non-hyperfinite,
  non-isomorphic factors,'' Commun.\ Pure Appl.\ Math.\ {\bf 16}, 19--26
  (1963).
  
\bibitem{srinivas} Srinivas, M., ``Collapse postulate for observables
  with continuous spectra,'' Commun.\ Math.\ Phys.\ {\bf 71}, 131--158
  (1980).
  
\bibitem{takesaki} Takesaki, M., {\it Theory of Operator Algebras} I
  (Springer, New York, 1979).
  
\bibitem{tomita} Tomita, M., ``Harmonic analysis on locally compact
  groups,'' Math.\ J.\ Okayama Univ.\ {\bf 5}, 133-193 (1956).
  
  

\end{thebibliography}
 \end{document}